# Coefficients of factor score determinacy for mean plausible values of Bayesian factor analysis

André Beauducel* & Norbert Hilger

*Institute of Psychology, University of Bonn, Germany*

## Abstract

In the context of Bayesian factor analysis, it is possible to compute mean plausible values, which might be used as covariates or predictors or in order to provide individual scores for the Bayesian latent variables. Previous simulation studies ascertained the validity of the plausible values by the mean squared difference of the plausible values and the generating factor scores. However, the generating factor scores are unknown in empirical studies so that an indicator that is solely based on model parameters is needed in order to evaluate the validity of factor score estimates in empirical studies. The coefficient of determinacy is based on model parameters and can be computed whenever Bayesian factor analysis is performed in empirical settings. Therefore, the central aim of the present simulation study was to compare the coefficient of determinacy based on model parameters with the correlation of mean plausible values with the generating factors. It was found that the coefficient of determinacy yields an acceptable estimate for the validity of mean plausible values. As for small sample sizes and a small salient loading size the coefficient of determinacy overestimates the validity, it is recommended to report the coefficient of determinacy together with a bias-correction in order to estimate the validity of mean plausible values in empirical settings.

Keywords: Bayesian factor analysis, plausible values, coefficient of determinacy

* Corresponding author:
André Beauducel, Institute of Psychology, Rheinische Friedrich-Wilhelms-Universität Bonn, Kaiser-Karl-Ring 9, 53111 Bonn, Germany. Email: beauducel@uni-bonn.de

Asparouhov and Muthén (2010a) proposed plausible values as indicators of individual scores on latent variables or factors that are computed in the context of Bayesian structural equation modeling (Muthén & Asparouhov, 2012). Although there has been some critique of Bayesian structural equation modeling (Stromeyer, Miller, Sriramachandramurthy, & DeMartino, 2015), several critical issues were successfully addressed by means of a series of specifications (Asparouhov, Muthén, & Morin, 2015). However, there are cautionary notes on the use of default settings for convergence (Zitzmann & Hecht, 2019) or on the use of default prior distributions (Smid & Winter, 2020). Nevertheless, Bayesian factor analysis (BFA) is meanwhile regularly used (e.g., Bonafede et al., 2021; Wang, Wang, Stein, Liu, & Chen, 2021). It follows from the regular use of BFA that plausible values as factor score estimates computed BFA (Asparouhov & Muthén, 2010a, b) may also be used regularly. Asparouhov and Muthén (2010a) have shown and discussed different applications and advantages of plausible value factor scores. However, the use of plausible values as indicators of the factors computed by means of BFA requires an indicator for the validity of plausible values.

As individual scores on factors can be of practical interest, it is not surprising that factor score estimates with different interesting properties have been proposed and evaluated in the framework of conventional factor analysis (Mulaik, 2010; Grice, 2001; McDonald, 1981, 2011). A well-known property of factor score estimates is their determinacy or validity, i.e., the correlation of the factor score estimates with the factor (Guttman, 1955; Beauducel & Hilger, 2015). Sometimes, when the factor score means are of relevance, the mean squared error (MSE), i.e., the mean squared difference between the factor scores and the true factor is also evaluated (Krijnen, Wansbeek, Ten Berge, 1996). The best mean-centered linear factor score estimate has the smallest MSE and the highest determinacy, i.e., the highest correlation with the factor. It has been recommended to report the coefficient of determinacy whenever factor score estimates are computed in empirical studies, where the original factor scores are principally unknown (Ferrando & Lorenzo-Seva, 2018; Grice, 2001). It has also been noted that the squared coefficient of determinacy may even be reported as an indicator of the reliability of factor score estimates (Ferrando & Lorenzo-Seva, 2018). Therefore, the aim of present study was to investigate how well the coefficient of determinacy and the MSE indicate the validity of the mean plausible values when computed from the Bayesian model parameters.

In the following, the coefficient of determinacy and the MSE as well as the evaluation of the validity of mean plausible values by means of these coefficients will be reviewed in more detail. After some definitions, it will first be shown that these coefficients are only based on model parameters so that they do not need the availability of the generating factor scores. Therefore, they can be used in any empirical study as indicators for the validity of plausible values. Second, in order to evaluate the precision of the coefficients based on model parameters, a simulation study is performed that compares the coefficients based on the generating factor scores with the coefficients based on model parameters.

## Definitions

The population common factor model can be defined as

$$\mathbf{x} = \mathbf{\Lambda f} + \mathbf{\Psi e}, \tag{1}$$

where $\mathbf{f}$ is a normally distributed random vector representing $q$ common factors, $\mathbf{\Lambda}$ is a $p \times q$ matrix of common factor loadings, where $p$ represents the number of observed variables, $\mathbf{e}$ is a normally distributed random vector of $p$ linear independent unique factors, and $\mathbf{\Psi}$ is a $p \times p$ positive definite matrix of unique or error factor loadings. It is assumed that $E(\mathbf{f}) = \mathbf{0}, E(\mathbf{ff}') = \mathbf{\Phi}, diag(\mathbf{\Phi}) = \mathbf{I}_q$ $E(\mathbf{e}) = \mathbf{0}, E(\mathbf{fe}') = \mathbf{0}$, and $E(\mathbf{ee}') = \mathbf{I}$, resulting in

$$\mathbf{\Sigma} = E(\mathbf{xx}') = \mathbf{\Lambda\Lambda}' + \mathbf{\Psi}^2. \tag{2}$$

It follows from the model assumptions, that post-multiplication of Equation (1) with the transpose of the generating factor yields

$$E(\mathbf{xf}') = \mathbf{\Lambda\Phi}, \tag{3}$$

so that the covariance of the generating factor with the measured variables can be expressed in terms of the model parameters $\mathbf{\Lambda}$ and $\mathbf{\Phi}$.

## Coefficients of determinacy and MSE

The best linear factor score estimate, i.e., the factor score estimate with the highest correlation with the factor, has already been proposed by Thurstone (1935) and can be written as

$$\hat{\mathbf{f}}_{BL} = \mathbf{\Phi\Lambda}'\mathbf{\Sigma}^{-1}\mathbf{x} = \mathbf{B}'_{BL}\mathbf{x}, \tag{4}$$

where $\mathbf{\Phi}$ represents the $q \times q$ matrix of factor inter-correlations, $\mathbf{\Lambda}$ the $p \times q$ matrix of factor loadings, $\mathbf{\Sigma}$ the $p \times p$ covariance matrix of observed variables, and $\mathbf{x}$, the random vector of $p$ observed variables, $\mathbf{B}_{BL}$ is the coefficient matrix for the best linear factor score estimate. The correlation of $\hat{\mathbf{f}}_{BL}$ with the factor $\mathbf{f}$ is the coefficient of determinacy (Ferrando & Lorenzo-Seva, 2018), which is given by

$$\begin{aligned} \mathbf{CD}_{BL} &= diag(E(\hat{\mathbf{f}}_{BL}\hat{\mathbf{f}}'_{BL}))^{-1/2} diag(E(\hat{\mathbf{f}}_{BL}\mathbf{f}')) = diag(\mathbf{\Phi\Lambda}'\mathbf{\Sigma}^{-1}\mathbf{\Lambda\Phi})^{-1/2} diag(E(\mathbf{\Phi\Lambda}'\mathbf{\Sigma}^{-1}\mathbf{xf}')) \\ &= diag(\mathbf{\Phi\Lambda}'\mathbf{\Sigma}^{-1}\mathbf{\Lambda\Phi})^{1/2}. \end{aligned} \tag{5}$$

Equation 5 is of interest when factor score estimates are computed in the context of empirical investigations because it is only based on model parameters because it is based on Equation 3. Therefore, Equation 5 allows to compute an indicator for the validity of factor score estimates even when the original factor scores are not available. Equation 5 follows from the population factor model. However, some amount of model misfit might occur because of sampling error, model misspecification, or model error of the factor model (MacCallum, 2003). As the amount of model fit depends on the estimation method (MacCallum, Browne, & Cai, 2007) and as the effect of Bayesian estimation on the coefficient of determinacy has until now not been investigated, it is not yet known how well Equation 5 represents the validity of mean plausible values. A related question is whether the mean plausible value factor scores $\hat{\mathbf{f}}_{PV}$ of a large number of imputations is a proxy of the best linear factor score estimate when the loadings $\mathbf{\Lambda}$ and factor inter-correlations $\mathbf{\Phi}$ are estimated from BFA.

If $\hat{\mathbf{f}}_{PV} \neq \hat{\mathbf{f}}_{BL}$, as might occur in the sample, the factor score coefficients for the mean plausible values $\mathbf{B}_{PV}$ can be computed *a posteriori* because $\hat{\mathbf{f}}_{PV} = \mathbf{B}'_{PV}\mathbf{x}$ implies

$$\mathbf{B}_{PV} = \mathbf{\Sigma}^{-1}\mathbf{x}\hat{\mathbf{f}}_{PV}. \tag{6}$$

The factor score determinacy for $\hat{\mathbf{f}}_{PV}$ is then

$$\mathbf{CD}_{PV} = diag(E(\hat{\mathbf{f}}_{PV}\hat{\mathbf{f}}'_{PV}))^{-1/2} diag(E(\hat{\mathbf{f}}_{PV}\mathbf{f}')) = diag(E(\hat{\mathbf{f}}_{PV}\hat{\mathbf{f}}'_{PV}))^{-1/2} diag(E(\hat{\mathbf{f}}_{PV}\mathbf{x}'\mathbf{\Sigma}^{-1}\mathbf{\Lambda}\mathbf{\Phi})). \tag{7}$$

Although one would expect very similar results for the computation of factor score determinacy according to Equations (5) and (7), the possible differences between these equations should be evaluated. Moreover, these two methods to compute the factor score determinacy should be compared with the correlation of the mean plausible values with the original factor scores. However, Luo and Dimitrov (2018) computed the correlation of the mean plausible values of Bayesian factor analysis with the generating factors, i.e. factor score determinacy, in the context of a simulation study. They found sufficiently high correlations of the mean plausible values with the original factors when the mean plausible values are based on at least 20 imputations. However, they did not compare the correlation of the mean plausible values with the factor scores with the coefficient of determinacy as it is computed from the model parameters by means of Equation (5) and (7).

Moreover, Budescu (1982) discussed the coefficient of determinacy in the context of multiple correlation and proposed to perform bias-correction similarly to the computation of the adjusted multiple correlation. Accordingly, Budescu (1982, p. 973, Eq. 3) proposed the following equation for the computation of bias for the squared coefficient of determinacy

$$\text{Bias}(\mathbf{CD^2}) = \left(\frac{p-2}{n-p-1}\right)\left(1-\mathbf{CD^2}\right) + \left(\frac{2(n-3)}{(n-p)^2-1}\right)\left(1-\mathbf{CD^2}\right)^2. \tag{8}$$

Accordingly, a bias-corrected coefficient of determinacy can be computed as $\mathbf{CD}_{corr.} = (\mathbf{CD^2} - \text{Bias}(\mathbf{CD^2}))^{1/2}$. Although the bias correction can be applied for the coefficients computed according Equation (5) and (7), it will be performed only for the conventional coefficient of determinacy because Budescu (1982) proposed this correction for this coefficient.

The MSE may provide different results than the coefficient of determinacy when the mean or standard deviation of the mean plausible values are different from the mean and standard deviation of the factor scores. Asparouhov and Muthén (2010a) compared the MSE of mean plausible values based on 500 imputations with the MSE of the best linear factor score estimate computed from maximum likelihood confirmatory factor analysis and found an extreme similarity for 10,000 cases. Luo and Dimitrov (2018) found that even fewer imputations may lead to an appropriate MSE for mean plausible values. As Asparouhov and Muthén (2010a) and Luo and Dimitrov (2018) used the difference between the mean plausible values and the factor scores in order to compute the MSE, they did not use an MSE that is based on model parameters.

Although the MSE can be computed from the difference between the mean plausible values and the original factor scores in the context of simulation studies, this is impossible in empirical studies. It is therefore of interest to compute the MSE in a way that does not require the original factor scores. When the MSE is computed from the model parameters and from mean centered $\hat{\mathbf{f}}_{BL}$, it follows from Krijnen et al. (1996, p. 3016, Equation 5) that it can be written as

$$\text{MSE}(\hat{\mathbf{f}}) = \left|(\boldsymbol{\Sigma}\mathbf{B} - \boldsymbol{\Lambda}\boldsymbol{\Phi})'\boldsymbol{\Sigma}^{-1}(\boldsymbol{\Sigma}\mathbf{B} - \boldsymbol{\Lambda}\boldsymbol{\Phi}))\right| + (\boldsymbol{\Phi} - \boldsymbol{\Phi}\boldsymbol{\Lambda}'\boldsymbol{\Sigma}^{-1}\boldsymbol{\Lambda}\boldsymbol{\Phi}). \tag{9}$$

As the matrix in braces is positive definite Krijnen et al. (1996) conclude that $\text{MSE}(\hat{\mathbf{f}})$ is minimal over all matrices $\mathbf{B}$ if $(\boldsymbol{\Sigma}\mathbf{B} - \boldsymbol{\Lambda}\boldsymbol{\Phi}) = \mathbf{0}$, which is given for $\mathbf{B} = \boldsymbol{\Sigma}^{-1}\boldsymbol{\Lambda}\boldsymbol{\Phi} = \mathbf{B}_{BL}$ so that

$$\text{MSE}(\hat{\mathbf{f}}_{BL}) = \boldsymbol{\Phi} - \boldsymbol{\Phi}\boldsymbol{\Lambda}'\boldsymbol{\Sigma}^{-1}\boldsymbol{\Lambda}\boldsymbol{\Phi} = \boldsymbol{\Phi} - \mathbf{C}\mathbf{D}_{BL}^2. \tag{10}$$

As the original factors are usually $z$-standardized, Equation 10 implies that for corresponding factors one can compute

$$diag(\text{MSE}(\hat{\mathbf{f}}_{BL})) = \mathbf{I} - \mathbf{C}\mathbf{D}_{BL}^2. \tag{11}$$

If $\hat{\mathbf{f}}_{PV} \neq \hat{\mathbf{f}}_{BL}$, as might occur in the sample, it follows from Krijnen et al. (1996, p. 3016, Eq. 4) that $\text{MSE}(\hat{\mathbf{f}}_{PV})$ for corresponding $\hat{\mathbf{f}}_{PV}$ and $\mathbf{f}$ can be computed as

$$diag(\text{MSE}(\hat{\mathbf{f}}_{PV})) = diag(E(\hat{\mathbf{f}}_{PV}\hat{\mathbf{f}}'_{PV}) - E(\boldsymbol{\Phi}\boldsymbol{\Lambda}'\boldsymbol{\Sigma}^{-1}\mathbf{x}\hat{\mathbf{f}}'_{PV}) - E(\hat{\mathbf{f}}_{PV}\mathbf{x}'\boldsymbol{\Sigma}^{-1}\boldsymbol{\Lambda}\boldsymbol{\Phi}) + \boldsymbol{\Phi}). \tag{12}$$

As for Equations (2) and (4), the similarity of the results from equations (11) and (12) depends on the similarity of $\hat{\mathbf{f}}_{BL}$ and $\hat{\mathbf{f}}_{PV}$, which should increase with sample size.

A simulation study was performed in order to ascertain whether the coefficient of determinacy can be computed from the conventional Equation (5) or whether the more specific Equation (7) should be preferred and whether the MSE can be computed from the conventional Equation (10) or whether the more specific Equation (11) should be preferred. As the generating factor scores are available in a simulation study, the results of Equations (5) and (7) can be compared with the correlation of $\hat{\mathbf{f}}_{PV}$ with $\mathbf{f}$ and the results of Equations (11) and (12) can be compared with the difference of $\hat{\mathbf{f}}_{PV}$ and $\mathbf{f}$. As a related issue, the correlation of $\hat{\mathbf{f}}_{PV}$ with $\hat{\mathbf{f}}_{BL}$ will also be computed.

## Simulation study

*Model specification and estimation*

Population models were based on $q \in \{3, 5\}$ common factors, low and high salient loadings $l \in \{.40, .80\}$, and six variables with salient loadings per factor $p/q = 6$. Moreover, all non-salient loadings were zero (independent clusters model), and we used population models with uncorrelated factors as well as population models with correlated common factors, $\rho \in \{.00, .40\}$. Small and large sample sizes were investigated, $n \in \{200, 800\}$. For each of these 2 ($q$) $\times$ 2 ($l$) $\times$ 2 ($\rho$) $\times$ 2 ($n$) = 16 conditions 1,000 samples of normally distributed observed variables computed from normally distributed common factors f~N(0,1) and error factors e~N(0,1) were drawn and submitted to BFA.

Data generation was performed with IBM SPSS, Version 26 and BFA was performed with Mplus 8.6. The salient loadings were freely estimated and the non-zero loadings were estimated with normally distributed priors with a zero mean and a variance of $\sigma^2=0.01$. The model specification was in line with the population models, i.e., the effects model misfit was not investigated. In consequence for each data set only a single BFA, based on freely estimated salient loadings, prior specification for non-salient loadings, and unit-variance of the factors, was performed. Zitzmann and Hecht (2019) emphasized the relevance of the precision of BFA estimates. They propose the effective sample size (ESS) as an indicator for the precision of parameter estimation. As Mplus 8.6 does not calculate ESS, but the potential scale reduction (PSR), the PSR has to be transformed into ESS. Here we use an ESS of 100 samples resulting in an PSR of 1.01 (Zitzmann & Hecht, 2019) and in a BCONVERGENCE option of 0.01, which is smaller than the Mplus default of 0.05. In order to ensure convergence, we used the BITERATIONS option in order to perform 100,000 iterations when necessary. The mean plausible values were computed from 1,000 imputations. An example for the Mplus syntax can be found in the Appendix.

We computed the following dependent variables in order to evaluate the coefficients of determinacy: The expected value of determinacy, i.e., correlation of the mean plausible values with the generating factor scores ($GDet_{PV}$), the coefficient of determinacy according to Equation 7 based on model parameters and parameters of the mean plausible values ($CD_{PV}$), the coefficient of determinacy according to Equation 5 based solely on model parameters ($CD_{BL}$), and expected value for the determinacy of the best linear predictor, i.e. the correlation of the best linear predictor computed from Equation 4 with the generating factor scores ($GDet_{BL}$). The mean squared difference of $GDet_{PV}$ and $CD_{PV}$ and the mean squared difference of $GDet_{BL}$ and was $CD_{BL}$ was calculated in order to evaluate the precision of the coefficients when compared to the expected determinacies computed from the generating factor scores. The bias-correction proposed by Budescu (1982) was performed on the means of $CD_{BL}$.

The following dependent variables were computed in order to evaluate the MSE of the mean plausible values: The expected value for the MSE, i.e., the mean squared difference of the mean plausible values and the generating factor scores ($GMSE_{PV}$), the MSE according to Equation 12 based on model parameters and parameters of the mean plausible values ($MSE_{PV}$), the MSE according to Equation 11 based solely on model parameters ($MSE_{BL}$), and the MSE as the mean squared difference between the best linear predictor computed from Equation 1 and the generating factor scores ($GMSE_{BL}$). The mean squared difference of expected value $GMSE_{PV}$ and the estimated $MSE_{PV}$ and the mean squared difference of the expected value $GMSE_{BL}$ and $MSE_{BL}$ was calculated in order to evaluate the precision of the estimated MSE.

*Results*

The precision of the coefficients of determinacy computed from plausible value parameters and model parameters ($CD_{PV}$, Equation 7) and solely from model parameters ($CD_{BL}$, Equation

5) was similar (see Table 1). Thus, there is no advantage of the slightly more complex computation by means of Equation 7 and the conventional computation according to Equation 5 can be regarded as sufficient. In line with this result, the mean correlation of $\hat{\mathbf{f}}_{PV}$ with $\hat{\mathbf{f}}_{BL}$ was greater than .99 in all conditions. It should be noted that the mean determinacy is substantially overestimated by $CD_{PV}$ as well as $CD_{BL}$ in the conditions based on small sample sizes and small salient loadings. Overall, the correction proposed by Budescu (1982) reduced the bias of $CD_{BL}$ considerably although it does not completely eliminate the bias for $q = 3$ (see Table 1).

Table 1

Expected values and mean squared errors of coefficients of factor score determinacy

| $\rho$ | $q$ | $l$ | $n$ | $GDet_{PV}$ | $CD_{PV}$ | $MSE(CD_{PV})$ | $GDet_{BL}$ | $CD_{BL}$ | $MSE(CD_{BL})$ | $CD_{BL,cor}$ |
|---|---|---|---|---|---|---|---|---|---|---|
| .0 | 3 | .4 | 200 | .707 | .755 | .004 | .708 | .748 | .003 | .719 |
| | | | 800 | .722 | .738 | .001 | .723 | .739 | .001 | .733 |
| | | .8 | 200 | .955 | .950 | .000 | .955 | .959 | .000 | .956 |
| | | | 800 | .956 | .953 | .000 | .956 | .957 | .000 | .957 |
| | 5 | .4 | 200 | .702 | .768 | .006 | .704 | .756 | .004 | .706 |
| | | | 800 | .719 | .743 | .001 | .720 | .745 | .001 | .734 |
| | | .8 | 200 | .955 | .948 | .000 | .955 | .961 | .000 | .955 |
| | | | 800 | .956 | .952 | .000 | .956 | .959 | .000 | .957 |
| .4 | 3 | .4 | 200 | .732 | .768 | .003 | .732 | .762 | .002 | .736 |
| | | | 800 | .745 | .755 | .000 | .745 | .757 | .000 | .750 |
| | | .8 | 200 | .956 | .952 | .000 | .956 | .960 | .000 | .956 |
| | | | 800 | .957 | .955 | .000 | .957 | .958 | .000 | .957 |
| | 5 | .4 | 200 | .738 | .782 | .003 | .738 | .772 | .003 | .726 |
| | | | 800 | .753 | .766 | .000 | .754 | .768 | .000 | .758 |
| | | .8 | 200 | .956 | .950 | .000 | .956 | .962 | .000 | .955 |
| | | | 800 | .957 | .954 | .000 | .957 | .959 | .000 | .958 |

*Note.* Population = independent clusters model with 6 salient loadings per factor, $q$ = number of factors, $l$ = salient loading size, $n$ = sample size; $\rho$ = factor inter-correlation; 1,000 samples per condition; $PV$ = mean plausible values of 1,000 imputations per sample; $BL$ = best linear factor score estimate; $GDet$ = correlation between estimates and true scores; $CD$ = coefficient of determinacy (cf. Eq. 5, Eq. 7); $MSE$ = mean squared error, $CD_{BL,cor}$ = bias-corrected, conventional coefficient of determinacy.

The results are similar for MSE, where precision of the computation from plausible values parameters and model parameters ($MSE_{PV}$, Equation 12) and the computation solely from model parameters ($MSE_{BL}$, Equation 11) was similar (see Table 2). Again, there is no advantage of the slightly more complex computation by means of Equation 12. The mean MSE is underestimated in the condition based on small sample sizes and small salient loadings.

Table 2

Expected values and mean squared errors of MSE estimates

| $\rho$ | $Q$ | $l$ | $n$ | $GMSE_{PV}$ | $MSE_{PV}$ | $MSE(MSE_{PV})$ | $GMSE_{BL}$ | $MSE_{BL}$ | $MSE(MSE_{BL})$ |
|---|---|---|---|---|---|---|---|---|---|
| .0 | 3 | .4 | 200 | .500 | .432 | .010 | .501 | .440 | .009 |
| | | | 800 | .479 | .456 | .002 | .478 | .453 | .002 |
| | | .8 | 200 | .095 | .098 | .000 | .094 | .080 | .000 |
| | | | 800 | .088 | .091 | .000 | .088 | .083 | .000 |
| | 5 | .4 | 200 | .508 | .416 | .014 | .511 | .427 | .013 |
| | | | 800 | .483 | .449 | .002 | .482 | .444 | .003 |
| | | .8 | 200 | .098 | .104 | .000 | .094 | .076 | .001 |
| | | | 800 | .089 | .095 | .000 | .088 | .081 | .000 |
| .4 | 3 | .4 | 200 | .464 | .412 | .007 | .466 | .418 | .007 |
| | | | 800 | .445 | .430 | .001 | .445 | .427 | .001 |
| | | .8 | 200 | .093 | .096 | .000 | .092 | .079 | .000 |
| | | | 800 | .086 | .088 | .000 | .086 | .082 | .000 |
| | 5 | .4 | 200 | .457 | .393 | .008 | .459 | .403 | .008 |
| | | | 800 | .434 | .414 | .001 | .432 | .410 | .001 |
| | | .8 | 200 | .095 | .100 | .000 | .092 | .075 | .000 |
| | | | 800 | .086 | .090 | .000 | .085 | .080 | .000 |

*Note.* Population = independent clusters model with 6 salient loadings per factor, $q$ = number of factors, $l$ = salient loading size, $n$ = sample size; $\rho$ = factor inter-correlation; 1,000 samples per condition; $PV$ = mean plausible values of 1,000 imputations per sample; $BL$ = best linear factor score estimate; $MSE$ = mean squared error; $MSE_{PV}$, $MSE_{BL}$ = estimated MSE of score estimates (cf. Eq. 11 and Eq. 12).

## Discussion

As plausible values have been proposed in the context of BFA (Asparouhov & Muthén, 2010a), they might be used as covariates or predictors in research settings or in order to provide individual scores for the BFA factors in applied settings. Accordingly, studies of the validity of plausible values are needed. Several studies on plausible values ascertained the validity of the scores by means of the correlation or squared difference of the plausible values and the generating factor scores (Asparouhov & Muthén, 2010a; Luo & Dimitrov, 2018). However, the generating factor scores are unknown in empirical studies so that the coefficient of determinacy has to be calculated in order to evaluate the validity of factor score estimates (Ferrando & Lorenzo-Seva, 2018). The coefficient of determinacy is solely based on model parameters and can be computed whenever BFA is performed in an empirical setting. Therefore, a central aim of the present study was to compare the coefficient of determinacy based on BFA model parameters with the correlation of mean plausible values with the generating factors. As plausible values are not computed in the same way as conventional factor scores (Asparouhov & Muthén, 2010a, 2010b), it was investigated whether an alternative coefficient of determinacy that is partly based on the mean plausible values yields an improved validity estimate.

It was found that both the conventional coefficient of determinacy and the plausible value-based coefficient of determinacy yield similar results. Moreover, the correlation of the mean plausible values with the best linear predictor computed from the BFA model parameters was close to one. This indicates that the conventional coefficient of determinacy, which is solely based on BFA model parameters yields an acceptable estimate for the validity of mean plausible values. It should be noted that for small sample sizes and a small salient loading size, both coefficients of determinacy overestimate the validity of mean plausible values. The correction proposed by Budescu (1982) reduced the bias of the conventional coefficient of determinacy considerably. It might therefore be recommended to use the conventional coefficient of determinacy together with the bias-correction in order to estimate the validity of mean plausible values. However, it is recommended to report the uncorrected as well as the bias-corrected values so that the amount of bias can also be considered.

Since the similarity of factor scores and plausible values has already been evaluated by means of the MSE, the MSE was also considered. It was shown that for mean centered best linear predictors the MSE can directly be computed from the conventional coefficient of determinacy. However, an MSE version based on the mean plausible values was also proposed. The results of the simulation study reveal that there was no substantial difference between the two versions of MSE so that the conventional MSE may be used. A bias was also found for small samples and small salient loadings resulting in an underestimation of MSE in this condition. One may consider to calculate the conventional MSE from the bias-corrected conventional coefficient of determinacy in order to reduce the bias of MSE.

As a limitation we note that the means in the population of the scores were zero which has enhanced the similarity of the coefficients of determinacy and the MSE. Effects of non-zero means remain to be investigated. A further limitation refers to the conditions of the simulation study. Effects of model misfit on the coefficient of determinacy remain to be investigated. However, we would recommend that mean plausible values should only be computed for BFA models with a very good fit to the data because computing scores representing factors that are not clearly represented by the data could lead to erroneous conclusions. Moreover, different numbers of variables with salient loadings per factor, cross-loadings, a larger number of factors, and different prior variances for the zero-loadings are aspects that remain to be investigated. It should, however, be noted that we followed the recommendation of Zitzmann and Hecht (2019) in enhancing the precision of BFA by means of a smaller convergence criterion and a larger number of iterations than the default options of Mplus 8.6. This resulted in a substantial increase of computation time.

Nevertheless, the tentative conclusion of the present study is that the conventional coefficient of determinacy which has been recommended as an indicator of the validity of factor score estimates (Ferrando & Lorenzo-Seva, 2018) can also be used as an indicator of the validity of mean plausible values for BFA models with a very good fit. A further conclusion is that the bias-correction proposed by Budescu (1982) may help to improve the precision of the coefficient of determinacy. These conclusions are of special importance for the use of BFA and mean plausible values in empirical studies when the generating factor scores are unknown.

## Appendix

**Example for Mplus input:**

```
DATA:       FILE = "sample.csv";
VARIABLE:     NAMES ARE x1-x18;
        USEVARIABLES ARE x1-x18;
ANALYSIS:     ESTIMATOR = BAYES;
        BITERATIONS = 100000;
        BCONVERGENCE = 0.01;
MODEL:      F1 by x1* x2-x18 (f1p1-f1p18);
        F2 by x1* x2-x18 (f2p1-f2p18);
        F3 by x1* x2-x18 (f3p1-f3p18);
        F1@1; F2@1; F3@1;
MODEL PRIORS: f1p7-f1p18~N(0,0.01);
        f2p1-f2p6~N(0,0.01);
        f2p13-f2p18~N(0,0.01);
        f3p1-f3p12~N(0,0.01);
OUTPUT:     STANDARDIZED (STDYX);
SAVEDATA:    FILE = BAY_PV.dat;
        SAVE = FSCORES (1000);
        RESULTS ARE BAY_RESULTS.dat;
```